\documentstyle[12pt]{article}
\begin{document}
\title{ Robertson-Schr\"odinger intelligent states for two-body Calogero model}
\author{{\bf M. Daoud} \\
\\
LPMC , Department of physics,\\ University Ibn Zohr\ , \ P.O.Box
28/S , Agadir\\ Morocco}
\maketitle

\begin{abstract}
 Using the Gazeau-Klauder and Klauder-Perelomov coherent states , we derive the Robertson-Schr\"odinger intelligent states for two-body Calogero quantum system.
\end{abstract}

\vfill
%\begin{flushright}
%Typeset using \LaTeX
%\end{flushright}
\newpage %\break

\section{Introduction}
Coherent states play an important role in quantum physics [1-3]. The
term "coherent" itself originates in the current language of
quantum optics (for instance, coherent radiation). It was
introduced by Glauber in 1960 for quantum oscillator (see[1-3]). These states
are closest to classical ones. Indeed, at the very beginning of
quantum theory, Schr\"odinger was interested by states which
restore the classical behaviour of the position operator for a
quantum oscillator and show that the coherent states mediate a
smooth transition from classical to quantum mechanics. The most
important properties of such states are : (i) normalization, (ii) continuity in the
labeling, (iii) over-completion and (iv) temporal stability. The properties (i)-(iv) are the minimal set of conditions of generalized coherent states [4-5]. They can be
defined following three equivalent ways:
($\bf{D}_{1}$) eigenvectors of the annihilation operator, ($\bf{D}_{2}$) Unitary action of the Weyl-Heisenberg group on the vacuum, and ($\bf{D}_{3}$)
 Minimization of Heisenberg inequality. \\
More recently, coherent states associated
with various exactly solvable quantum mechanics have been
successfully studied [5-11] and was constructed following three unequivalents prescriptions  which extend the definitions ($\bf{D}_{1}$), ($\bf{D}_{2}$) and ($\bf{D}_{3}$). The coherent states obtained in the ($\bf{D}_{1}$) scheme
are called Gazeau-Klauder ones [5-7]. Using the ($\bf{D}_{2}$) definition,
one obtain coherent states of Klauder-Perelomov kind [8-10]. The extension of third
definition ($\bf{D}_{3}$) for quantum system other than harmonic
oscillator leads to the minimization of the
Robertson-Schr\"odinger uncertainty relation [12-13] (which generalizes
the usual Heisenberg inequality). In this case, one obtain the
so-called Robertson-Schr\"odinger or intelligent states [14-15]. In the
present work, we are interested by the construction of intelligent
states for the two-body-Calogero model. We give the explicit
analytical representation of  states which minimize the
Robertson-Schr\"odinger uncertainty relation. Two equivalents
representations are considered. The first one is related  the
Gazeau-Klauder coherent states and the second involves the states
constructed \`a la Klauder-Perelomov. Note that, for shortness
reasons, we will avoid the presentation of computation techniques
which are more or less similar to ones used in the derivation of
Intelligent states for other quantum systems(see [7-11])
 This letter is organized as
follows: The section 2 is devoted to a brief review of the two
body Calogero system in order the fix  the notations used through
this work. The third section concern the algebraic derivation of
Calogero states minimizing the Robertson-Schr\"odinger uncertainty
relation. The analytical solutions of the Intelligent states, for
the quantum system under consideration, are given in section 4. Concluding remarks
close this letter.
\section{\bf The two-body Calogero model}

 The one dimensional harmonic oscillator
\begin{equation}
H_{0}= -{1\over2}{d^2\over{dx^2}} + {1\over2}x^2
\end{equation}
still integrable if we add a $x^{-2}$ potential:
\begin{equation}
H_{cal} = a^+a^- + {1\over2}+  {{\eta}^{2}\over {x^2}}
\end{equation}
where
\begin{equation}
a^{\pm} = {1\over\sqrt{2}}(x\pm {d\over{dx}})
\end{equation}
are the usual harmonic creation and annihilation operators. The normalized eigenfunctions of the Hamiltonian $H_{cal}$ are
\begin{equation}
\Psi_n(x) = (-)^{n} \sqrt{{2n!}\over{\Gamma(n+e_{0})}}
L^{e_{0}-1}_n (x^{2})\exp(-{x^{2}\over{2}})
\end{equation}
where $L^{e_{0}-1}_n$ are the Laguerre polynomials and $e_{0} =
1 + \sqrt{({1\over4}+2{{\eta}^{2}})}$. The eigenvalues are
given by:
\begin{equation}
H_{cal} \Psi_n(x) = (2n+e_{0})\Psi_n(x)
\end{equation}
The waves functions $\Psi_n(x)$ form a basis in the Hilbert space $\cal H$ of square integrable functions on the half axis $0<x<\infty$. The raising and lowering operators are
defined by
\begin{equation}
A^{\pm} = {1\over2}((a^{\pm})^{2}-{{\eta}^{2}\over {x^2}}),
\end{equation}
and act on the eigenstates $\vert\Psi_n\rangle$  as follows
\begin{equation}
A^{+}\vert\Psi_n\rangle = \sqrt{(n+1)(n+1+e_{0})}e^{-i\beta f(n+1)} \vert\Psi_{n+1}\rangle{\hskip 2cm}
\end{equation}
and
\begin{equation}
A^{-}\vert\Psi_n\rangle = \sqrt{n(n+e_{0})} e^{+i\beta f(n)}\vert\Psi_{n-1}\rangle.{\hskip 2cm}
\end{equation}
The functions $f(n)$ are assumed to satisfy the following
condition
\begin{equation}
f(n) + f(n-1)+...+ f(1)= e_{n}
\end{equation}
which gives $f(1)=e_{0}+2$ and $f(n \ne 1)=2$. Notice that the value 2
is exactly the spacing between two successive energy levels. The
role of real parameter $\beta$ will be clarified, hereafter, when
we will discuss the temporal stability of the coherent states. Note also that the state $\Psi_n(x)$ can be obtained by applying the operator $(A^{+})^{n}$ on the
ground state $\Psi_0(x)$.

\section{Robertson-Schr\"odinger intelligent states:\\
Algebraic derivation}

\hspace*{0.5cm}We define two hermitains operators $A$ and $B$, in terms of the operators $A^{+}$ and $A^{-}$, as follows
\begin{equation}
A = {1\over\sqrt{2}}(A^{+}+A^{-})
\end{equation}
and
\begin{equation}
B = {1\over\sqrt{2}}(A^{+}-A^{-})
\end{equation}
which satisfy the commutation relation
\begin{equation}
\lbrack A,B\rbrack = i \lbrack A^{+},A^{-} \rbrack = iH_{cal}
\end{equation}
where $H_{cal}$ is the Hamiltonian of the Calogero system. It is well known that for two hermitian operators satisfying the non canonical commutation relations(12), the covariances $({\Delta}A)^2$ and $({\Delta}B)^2$ satisfy the Robertson-Schrodinger uncertainty relation\\
\begin{equation}
({\Delta}A)^2({\Delta}B)^2\ge{1\over4}({\langle H \rangle}^{2}+{\langle F \rangle}^{2})
\end{equation}
where the mean value of the operator $F$ is defined by
\begin{equation}
\langle F \rangle  = i ( ({\Delta}A^{+})^{2}- ({\Delta}A^{-})^{2})
\end{equation}
by means of the variances of the operators $A^{+}$ and $A^{-}$.\\
The so-called Robertson-Schr\"odinger intelligent states are obtained when the equality in the uncertainty relation (13) are realizead [12,13,14]. They satisfy the eigenvalues equation [14]
\begin{equation}
((1-{\lambda})A^{+}+(1+{\lambda})A^{-})\vert z,\beta, \lambda \rangle = 2z\vert z, \beta, \lambda \rangle
\end{equation}
For the states solutions of the latter equation , the variances of $A$ and $B$ are
\begin{equation}
({\Delta}A)^2  =  \frac{\vert\lambda\vert}  {2}\sqrt{\langle H\rangle ^{2}+\langle F\rangle ^{2}},
\end{equation}
and
\begin{equation}
({\Delta}B)^2  = \frac{1}{2\vert\lambda\vert} \sqrt{\langle H\rangle ^{2}+\langle F\rangle ^{2}}.
\end{equation}
Clearly, for $\vert\lambda\vert =1$, we have $({\Delta}A)^2=({\Delta}B)^2$ and the states are called, in this
case, the generalized coherent states. Expanding the states $\vert z,\beta, \lambda\rangle$ in the $\vert\Psi_{n}\rangle$ basis and using equations (7) and (8), one obtain the solutions of (15) for $\lambda \ne -1$. In compact form, they are given by (up to the normalization constant)
\begin{equation}
\vert z, \beta, \lambda \rangle\ = U(z,\lambda)\vert\Psi_{0}\rangle
\end{equation}
as the action of the operator
\begin{equation}
U(z,\lambda)= \sum_{n=0}^{\infty}{\bigg[
\bigg({2z\over1+{\lambda}}\bigg)
{1\over{A^{+}A^{-}}}A^{+}+\bigg({1-{\lambda}\over1+{\lambda}}\bigg){1\over{A^{+}A^{-}}}
(A^{+})^{2}\bigg]} ^{n}
\end{equation}
on the ground state $\vert\Psi_0\rangle$ of the Calogero quantum system. Note that we have used similar computation techniques to ones used in obtaining the intelligent states for other exactly solvable quantum mechanical systems like for instance the P\"oschl-Teller system [11]. The states (Eq.(18)), obtained by minimizing the Robertson-Schr\"odinger uncertainty relation, generalize the so-called Barut-Girardello coherent states. The latters , defined as the eigenvectors of the lowering operator $A^{-}$,  can be obtained by simply setting $\lambda = 1$ in (18) and (19).Their explicit expressions , in the
$\{\vert\Psi_{n}\rangle\}$ basis, and their properties will be given in the next section.
\section{\bf Robertson-Schr\"odinger Intelligent states:\\Analytical representations}
Using the analytical representations of the Barut-Girardello and  Klauder-Peremolov coherent states, we will construct the analytical representations of the Robertson-Schr\"odinger intelligent states. As we will see, the use of the analytical representation leads easily to the solutions of the eigenvalue equation (15). We start by giving the generalized intelligent states in the Barut-Girardello scheme.
\subsection{\bf Barut-Girardello analytic representation}
The Barut-Girardello coherent states of two-body Calogero system are defined as eigenstates of the lowering operator $A^{-}$
\begin{equation}
A^{-}\vert z,\beta \rangle\  =  z \vert z, \beta \rangle\
\end{equation}
Expanding the states $ \vert z,\beta \rangle\ $ in the basis
$\{\vert\Psi_{n}\rangle\}$ and using the action of the operator
$A^{-}$, one obtain
\begin{equation}
\vert z \rangle\  =  N(\vert z \vert)
\sum_{n=0}^{\infty}{z^{n}e^{-i\beta
e_{n}}\over{\sqrt{\Gamma(n+1)\Gamma(n+1+e_{0})}}}\vert
\Psi_{n}\rangle
\end{equation}
where the normalization constant $N(\vert z \vert)$ takes the form
\begin{equation}
N^{2}(\vert z \vert) ={{\vert z \vert}^{e_{0}}\over{I_{e_{0}}}(2
\vert z \vert)}
\end{equation}
and $I_{e_{0}}(2 \vert z \vert)$ is the modified Bessel
function. Remark that states (18) coincides with ones expressed by (21) for $\lambda = 1$. The two-body Calogero system coherent states solve the
unity
\begin{equation}
\int| z, \beta\rangle\langle z,\beta| = I_{\cal H}
\end{equation}
with respect the measure (computed using the inverse Mellin transform [16])
\begin{equation}
d\mu(z)=
\frac{2}{\pi}\ I_{e_{0}}(2r)K_{\frac{e_{0}}{2}}(2r)rdrd\phi
\end{equation}
where $z = r e^{i\phi}$. The coherent states (22) are stable
\begin{equation}
e^{-i H_{cal}}| z, \beta\rangle = | z, \beta + t \rangle
\end{equation}
This property is ensured by the presence of the phase factor $\beta$ in the definition of the actions of creation and annihilation operators (eqs.(7) and (8)). Due to the completion
property (23), any state $\vert f \rangle$, of the Hilbert space, can
be represented by an entire function
\begin{equation}
f(z,\beta) = \sqrt{I_{e_{0}}(2 \vert z \vert)\over{{\vert z
\vert}^{e_{0}}}}\langle \bar z , \beta \vert f \rangle
\end{equation}
In particular, the analytical representations of the vectors $\vert \Psi_{n}\rangle$ are
\begin{equation}
{\cal F}_{n}(z,\beta)  =  \frac{z^{n}e^{-i\beta
e_{n}}}{{\sqrt{\Gamma(n+1)\Gamma(n+1+e_{0})}}}
\end{equation}
The Ladder operators $A^{+}$ and $A^{-}$ are realized, in this representation, by
\begin{equation}
A^{+} = z
\end{equation}
\begin{equation}
A^{-} = z \frac{d^2}{dz^2} + e_0 \frac{d}{dz}
\end{equation}
This differential realization is an usefull tool to determine the
Robertson-Schr\"odinger intelligent states of the system under
consideration. Indeed, introducing the analytic function
\begin{equation}
\Phi_{(z',\lambda)}(z)  = \sqrt{I_{e_{0}}(2 \vert z
\vert)\over{{\vert z \vert}^{e_{0}}}}\langle \bar z , \beta \vert
z',\lambda, \beta \rangle,
\end{equation}
the eigenvalues equation(15) is converted into the following second order differential equation
\begin{equation}
\big[(1+{\lambda})(z {d^{2}\over{dz^{2}}} + e_{0}{d\over{dz}}) +
(1-{\lambda})z - 2z'\big]\Phi_{(z',\lambda)}(z)= 0.
\end{equation}
For $\lambda\ne\pm1$, the admissible solutions are given by
\begin{equation}
\Phi_{(z',\lambda)}(z)=
\exp\bigg(\pm\sqrt{\lambda-1\over\lambda+1}z\bigg){
_{1}F_{1}\big({e_{0}\over2}\pm z',
e_{0},\mp\sqrt{\lambda-1\over\lambda+1}z\big)}
\end{equation}
The upper and lower signs in Eq.(32) are equivalents because the
confluent hypergeometric function $_{1}F_{1}$ can be written in
two equivalent forms thanks to Kummer's transformation [16].
Due to the analytical properties of the hypergeometric functions, the squeezing parameter $\lambda$ satisfies the condition

\begin{equation}
Re(\lambda) > 0
\Longleftrightarrow
\sqrt{\vert\lambda-1\vert\over{\vert\lambda+1}\vert}< 1
\end{equation}

\subsection{\bf Klauder-Perelomov analytic representation}
The set of Klauder-Perelomov coherent states is obtained from the
lowest state $\vert\Psi_{0}\rangle$ as follows
\begin{equation}
\vert z,\beta \rangle = \exp(zA^{+}-\bar z
A^{-})\vert\Psi_{0}\rangle = D(z)\vert\Psi_{0}\rangle
\end{equation}
which can be expanded as
\begin{equation}
\vert z,\beta \rangle = \sum_{n=0}^{\infty}\langle\Psi_{n}\vert
D(z)\vert\Psi_{0}\rangle \vert\Psi_{n}\rangle
\end{equation}
To compute the matrix element occurring in the last equation, we
used the actions of operators $A^{+}$ and $A^{-}$ on the
states $\vert\Psi_{n}\rangle$ given by Eqs.(7)and (8). As result, we
have
\begin{equation}
\langle\Psi_{n}\vert D(z)\vert\Psi_{0}\rangle =
\sqrt{\Gamma(e_{0}+n)\over\ {\Gamma(e_{0})\Gamma(n+1)}}
{\kappa^{n}\over{({1+\vert\kappa\vert}^{2})}^{e_{0}/2+n/2}}e^{-(i\beta
e_{n})}
\end{equation}
where the variables $\kappa$ and $z$ are such that $\kappa$ =
$z\sinh(\vert z\vert)\over\vert z\vert$.To get the identity
resolution, it is more convenient to introduce the variable
$\zeta$=$\kappa\over {(1+{\vert\kappa\vert}^{2})}^{1/2}$. The
coherent states labeled now by the new variable $\zeta$ solve the
identity
\begin{equation}
\int d\mu(\zeta)\vert\zeta,\beta\rangle\langle\zeta,\beta\vert =
{\cal I}_{\cal H}
\end{equation}
where the measure is given by
\begin{equation}
d\mu(\zeta) = {e_{0}- 1\over\pi}
{d^{2}\zeta\over{({1-{\vert\zeta\vert}^{2}})}^{2}}
\end{equation}
Then, any state $\vert\Psi_{n}\rangle$ is represented by the
analytic function
\begin{equation}
{\cal G}_{n}(\zeta,\beta) = {\zeta^{n}}{\Gamma(e_{0}+n)\over\
{\Gamma(e_{0})\Gamma(n+1)}} e^{-i\beta e_{n}}
\end{equation}
The operators $A^{+}$ and $A^{-}$ act in the space of analytical
functions as first-order differential operators
\begin{equation}
A^{+} = \zeta^{2} \frac{d^2}{d\zeta^{2}} + e_{0}\zeta
\end{equation}
\begin{equation}
A^{-} ={ d\over{d\zeta}}
\end{equation}
Remark that the representations ${\cal F}_{n}(z,\beta)$ and ${\cal
G}_{n}(\zeta,\beta)$ are related via a transformation of Laplace
type
\begin{equation}
{\cal G}_{n}(\zeta,\beta)=
{\zeta^{-e_{0}}\over{\sqrt{\Gamma(e_{0})}}}\int_{0}^{+\infty}z^{e_{0}-1}{\cal
F}_{n}(z,\beta)\exp(-z/\zeta)dz
\end{equation}
At this stage, we have the necessary tools to obtain the
Robertson-Shr\"odinger states in the Klauder-Peremolov
representation. Indeed, defining the following function
\begin{equation}
\Phi_{(\zeta',\lambda)}(\zeta)={(1-{\vert\zeta\vert}^{2})^{-e_{0}/2}}\langle\bar{\zeta},\beta\vert
\zeta',\lambda,\beta\rangle,
\end{equation}
the eigenvalue equation (15) becomes
\begin{equation}
\bigg[[(1-\lambda)\zeta^{2}+
(1+\lambda)]\frac{d}{d\zeta}+(1+\lambda)e_{0}\zeta -
2\zeta'\bigg]\Phi_{(\zeta',\lambda)}(\zeta)= 0.
\end{equation}
The functions $\Phi_{(\zeta',\lambda)}(\zeta)$, satisfying equation (44), should be analytic
in the unit disk ( $\vert\zeta\vert<1$) [16] and are given by
\begin{equation}
\Phi_{(\zeta',\lambda)}(\zeta)= {\cal
N}^{-1/2}\bigg(1+\sqrt{\frac{\lambda-1}{\lambda+1}}\zeta\bigg)^{-e_{0}/2+\zeta'
\over{\sqrt{{\lambda}^{2}-1}}}
\bigg(1-\sqrt{\frac{\lambda-1}{\lambda+1}}\zeta\bigg)^{-e_{0}/2-\zeta'\over{\sqrt{{\lambda}^{2}-1}}}\hskip 0.5cm
\end{equation}
where ${\cal N}$ is a normalization constant and $Re(\lambda)>0$ (
a condition imposed, here again, by the analytical properties of the obtained
solutions). The states (32) and (45) are equivalents by virtue of Laplace transformation (42) relating Gazeau-Klauder and Klauder-Perelomov analytic representations.
\section{\bf Conclusion}
In this short note, we gave the intelligent states which minimize Schr\"odinger-Robertson uncertainty relation for two
body Calogero system. We constructed the Gazeau-Klauder coherent states, defined as the eigenstates of the lowering operator $A^-$. We proved that they are normalizable , continuous, constitute an overcomplete set and are temporally stable. We introduced also the coherent states constructed following the group theoretical approach ( Klauder-Perelomov method).
It is shown that the states defined $\grave{a}$ la Klauder-Perelomov satisfy Gazeau-Klauder minimal set of conditions
for generalized coherent states (i.e; noramlization, continuity in the labeling, resolution to identity and temporal stability). We established that the analytical representations of coherent sates allow the derivation of intelligent
states in an easy way. Other quantum systems, particularly those with continous energy spectrum, are under investigation and we hope to report on them in the near future.

\end{document}